\documentclass[amsmath,fixfloats,twocolumn]{revtex4}
\usepackage[pdftex]{graphicx}
\usepackage[pdftex,plainpages=false]{hyperref}

\newlength{\defbaselineskip}
\newcommand{\1}{{\mbox{\hspace{1pt}}}}

\newcommand{\eqn}[1]{Eq.~(\ref{#1})}
\newcommand{\Eqn}[1]{Equation~(\ref{#1})}
\newcommand{\fig}[1]{Fig.~\ref{#1}}
\newcommand{\Fig}[1]{Figure~\ref{#1}}
\newcommand{\secn}[1]{Section~\ref{#1}}
\newcommand{\Secn}[1]{Section~\ref{#1}}

\newcommand{\w}{\omega}
\newcommand{\scr}[1]{{\scriptscriptstyle{#1}}}

\newcommand{\mr}[1]{{\mathrm{#1}}}

\newcommand{\degrees}{\mbox{$^\circ$}}

\newcommand{\chem}[1]{$\mathrm{#1}$}

\newcommand{\ip}[2]{\left\langle{#1}|{#2}\right\rangle}

\begin{document}

\title{Effects of mode degeneracy in the LIGO Livingston Observatory recycling
cavity}
\author{Andri M. Gretarsson\footnote{greta9a1@erau.edu.  Comments may be directed to
either of the primary authors: E. D'Ambrosio and A.M. Gretarsson.}}
\affiliation{Embry-Riddle Aeronautical University, 3700 Willow Creek Rd., Prescott, AZ
86301, USA.}
\author{Erika D'Ambrosio\footnote{ambrosio@mail.jpl.nasa.gov}}
\affiliation{LIGO Laboratory, California Institute of Technology, MS 18-34, Pasadena, CA
91125, USA}
\author{Valery Frolov}
\affiliation{LIGO Livingston Observatory, Livingston LA 70754, USA.}
\author{Brian O'Reilly}
\affiliation{LIGO Livingston Observatory, Livingston LA 70754, USA.}
\author{Peter K. Fritschel}
\affiliation{LIGO Project, Massachusetts Institute of Technology, NW17-161,  175 Albany
Street, Cambridge, MA 02139, USA \vspace{3ex}}


\renewcommand{\arraystretch}{1.8}
\begin{abstract}
We analyze the electromagnetic fields in a Pound-Drever-Hall locked, marginally unstable,
Fabry-Perot cavity as a function of small changes in the cavity length during resonance.
More specifically, we compare the results of a detailed numerical model with the behavior
of the recycling cavity of the Laser Interferometer Gravitational-wave Observatory (LIGO)
detector that is located in Livingston, Louisiana. In the interferometer's normal mode of
operation, the recycling cavity is stabilized by inducing a thermal lens in the cavity
mirrors with an external \chem{CO_2} laser. During the study described here, this thermal
compensation system was not operating, causing the cavity to be marginally optically
unstable and cavity modes to become degenerate. In contrast to stable optical cavities,
the modal content of the resonating beam in the uncompensated recycling cavity is
significantly altered by very small cavity length changes. This modifies the error
signals used to control the cavity length in such a way that the zero crossing point is
no longer the point of maximum power in the cavity nor is it the point where the input
beam mode in the cavity is maximized.

\end{abstract}

\maketitle

\section{Introduction}\label{Introduction}

\begin{figure}
    \begin{center}
    \includegraphics[width=3.2in]{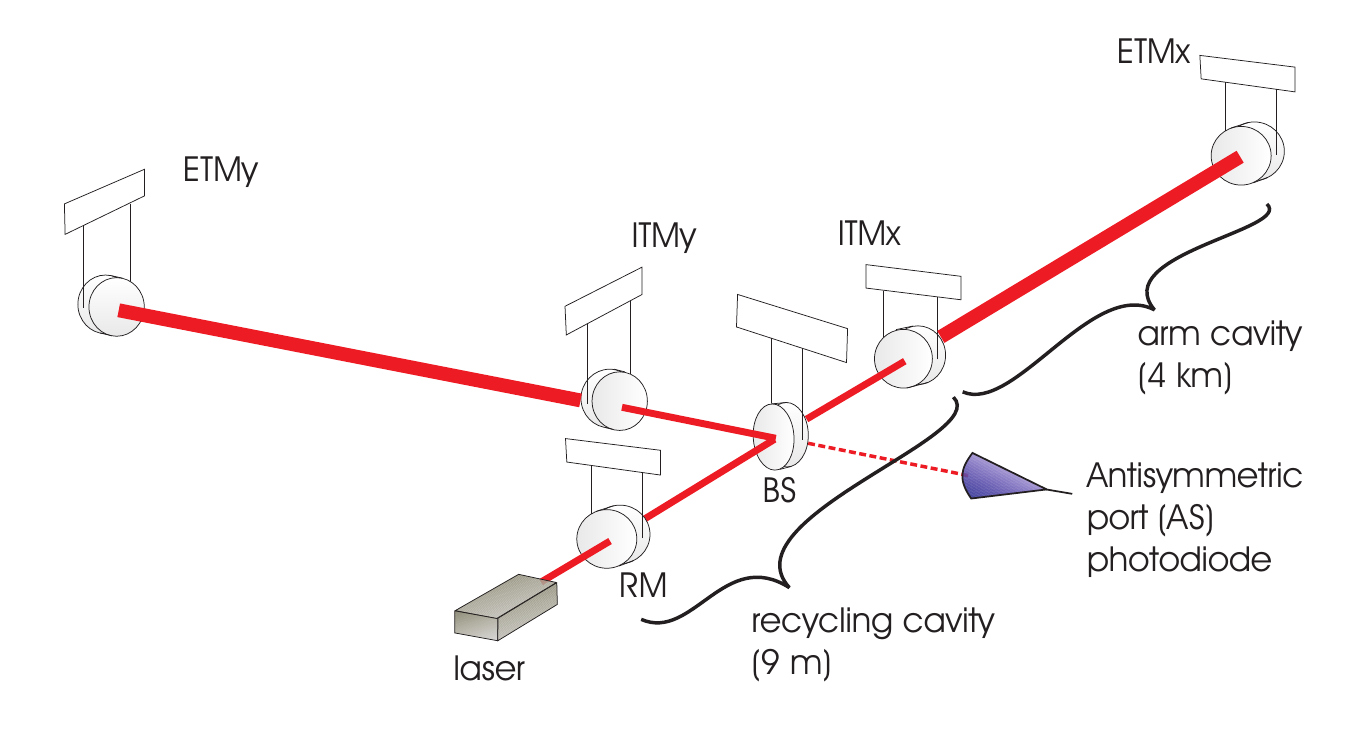}
    \caption{Arrangement of the core optics of the LIGO interferometers (not to scale).
    The recycling cavity is the cavity formed by the recycling mirror (RM), beamsplitter
    (BS), and the two input mirrors to the arm cavities (ITMx and ITMy).
    The recycling cavity length is the mean of the two optical
    paths between the recycling mirror (RM) and the input mirrors
    to the two arms (ITMx and ITMy). Note that since the interferometer is operated on a
    dark fringe, almost no carrier light escapes out the antisymmetric port of the
    beamsplitter to the photodiode at right.}
    \label{LIGOifo}
    \end{center}
    \end{figure}

The Laser Interferometer Gravitational Wave Observatory
(LIGO)~\cite{Abramovici1992,Fritschel2001b,Abbott2004} is a set of three kilometer-scale
suspended interferometers for the detection of gravitational waves from astronomical
sources~\cite{Weiss1999,Hough2005}.  Each of the three detectors consists of coupled
optical cavities with the basic arrangement of a Michelson interferometer as shown in
\fig{LIGOifo}. In this paper we are primarily concerned with the behavior of the
recycling cavity. The LIGO recycling cavity was designed to be optically stable at full
input power, which implies thermally induced lensing of the recycling cavity optics
during full power operation~\cite{Beusoleil2003,Kells1997}. In addition to the thermal
lens produced by the resonating beams at full power, active tuning of the input test mass
effective radii of curvature is achieved by a thermal compensation
system~\cite{TCStechdoc} which uses a \chem{CO_2} laser to heat the surface of the test
ITM's, creating a thermally induced lens.  Without the thermal compensation system, and
especially at low input power, the LLO recycling cavity is optically unstable. The
thermal compensation system was installed because the instability made proper
interferometer operation difficult. In addition to the effects described in this paper,
the marginal instability of the recycling cavity lead to reduced optical gain and
extremely high sensitivity to alignment fluctuations. This lead to greater than expected
sensitivity to seismic motion and made the wavefront sensor alignment
system~\cite{WFSoverview, BiplapNote} very difficult to operate.

In this paper, we present measurements of the fields in the recycling cavity when the
thermal compensation system is not operating and input power is set low enough so that no
significant thermal lensing occurs due to absorption of the resonating beam in the cavity
optics. In other words, we were studying the unstable cavity behavior. The measurements
are then compared to a detailed numerical model. In the absence of thermal lensing, the
recycling cavity is optically unstable with $g_1g_2\approx1.0004$ which takes into
account the curvature of the input test masses, the recycling mirror and also the slight
curvature of the beamsplitter. Therefore, the recycling cavity beam is not matched to the
fundamental optical mode of the stable interferometer arm cavities, nor to the input beam
which is designed to closely match the fundamental mode of the arms. In this ``cold''
condition, the optical fields that resonate only in the recycling cavity exhibit a
ringlike structure. The spatial structure of these fields is very sensitive to small
angle or length perturbations of the cavity. Basically, this occurs because marginally
unstable cavities have non-separable boundary conditions on the optical fields, leading
to mode degeneracy. So, the cavity will resonate arbitrary TEM fields as long as the
optical loss is small. To model the detailed behavior, it is no longer efficient to
express the cavity field in terms of the usual Hermite-Gaussian (or Laguerre-Gaussian)
modes~\cite{D'Ambrosio2006}. We therefore chose to compare our measurements to a
numerical simulation based on the Fast Fourier Transform as opposed to a simulation based
on the propagation of cavity modes. The simulation software used is referred to as "The
FFT Model"~\cite{Bochner1998}. As a result of the sensitivity of the transverse field
distribution to small length changes, we find that the Pound-Drever-Hall error
signal~\cite{Drever1983} used to control the length of the cavity is modified and the
lock-point develops an unexpected offset.

We tested the model against actual interferometer behavior by applying a series of
offsets to the error point of the length loop of the recycling cavity of the Livingston
interferometer~\cite{LLOwebsite}. The length loop of the recycling cavity controls the
position of the recycling mirror.

In this paper, the phrase ``cavity length'' always refers to the common arm length of the
cavity, that is the sum of the two optical paths in the cavity, to ITMx and ITMy, divided
by two (the quantity $L=(L_1+L_2)/2$ in \Fig{notation}).  By contrast, the differential
arm length is the difference of the two optical pathlengths divided by two. In order to
lock the recycling cavity, we need to control both the common arm length of the cavity
and the differential arm length of the cavity. The error signal for the common arm length
is generated at the reflected port using the light returning from the recycling mirror.
The differential length is sensed at the antisymmetric port using the very small amount
of sideband light that manages to enter the cavity, despite being non-resonant, and leaks
out of cavity via the beamsplitter. The differential armlength is controlled by
simultaneously actuating on the beamsplitter and recycling mirror in such a way that a
pure differential arm length change is achieved. The common arm length, or just ``the
cavity length'' is controlled by actuating on the recycling mirror alone.

We recorded the power in the recycling cavity for a range of cavity length loop error
point offsets and also acquired images of the light distribution inside the cavity and on
reflection from the cavity. We then compared our results to those of the simulation.
Although the FFT Model has previously been used to gain insight into power build-up
effects of the LIGO interferometers~\cite{DAmbrosio2004B}, this study is unusual in that
it provides a direct, well controlled, quantitative comparison between the predictions of
the model and the observed interferometer behavior.

\section{Measurement Overview}
\label{MeasurementOverview} The light entering the interferometer consists of carrier
light (the main laser frequency) and upper and lower phase-modulation sidebands with
modulation index $\Gamma\approx0.34$. When the interferometer is operating in its normal,
fully locked configuration, the carrier resonates in both the recycling cavity and in the
arm cavities, while the sidebands resonate only in the recycling cavity. To make the
optical fields easier to model and to make the results easier to interpret, most of our
measurements were taken when the arms were unlocked and the recycling cavity was locked
so that only the carrier was resonant in the recycling cavity while the sidebands were
non-resonant. By adding an offset to the error point of the Pound-Drever-Hall locking
servo controlling the position of the recycling mirror, we induced small length changes
of the cavity with respect to the null point of the servo. We could change the cavity
length by several nanometers before lock was lost. The length offset was calibrated in
terms of actual cavity length change in nanometers~\cite{andri2004} which allowed a
quantitative comparison between our simulation results and the experimental data. We
obtained the cavity power as a function of the length offset and obtained images of the
cavity beam profile at various offsets. We also made preliminary measurements with the
full interferometer locked.  No attempt was made to model the fields in the fully locked
interferometer.

\begin{figure}
    \begin{center}
    \includegraphics[width=3.2in]{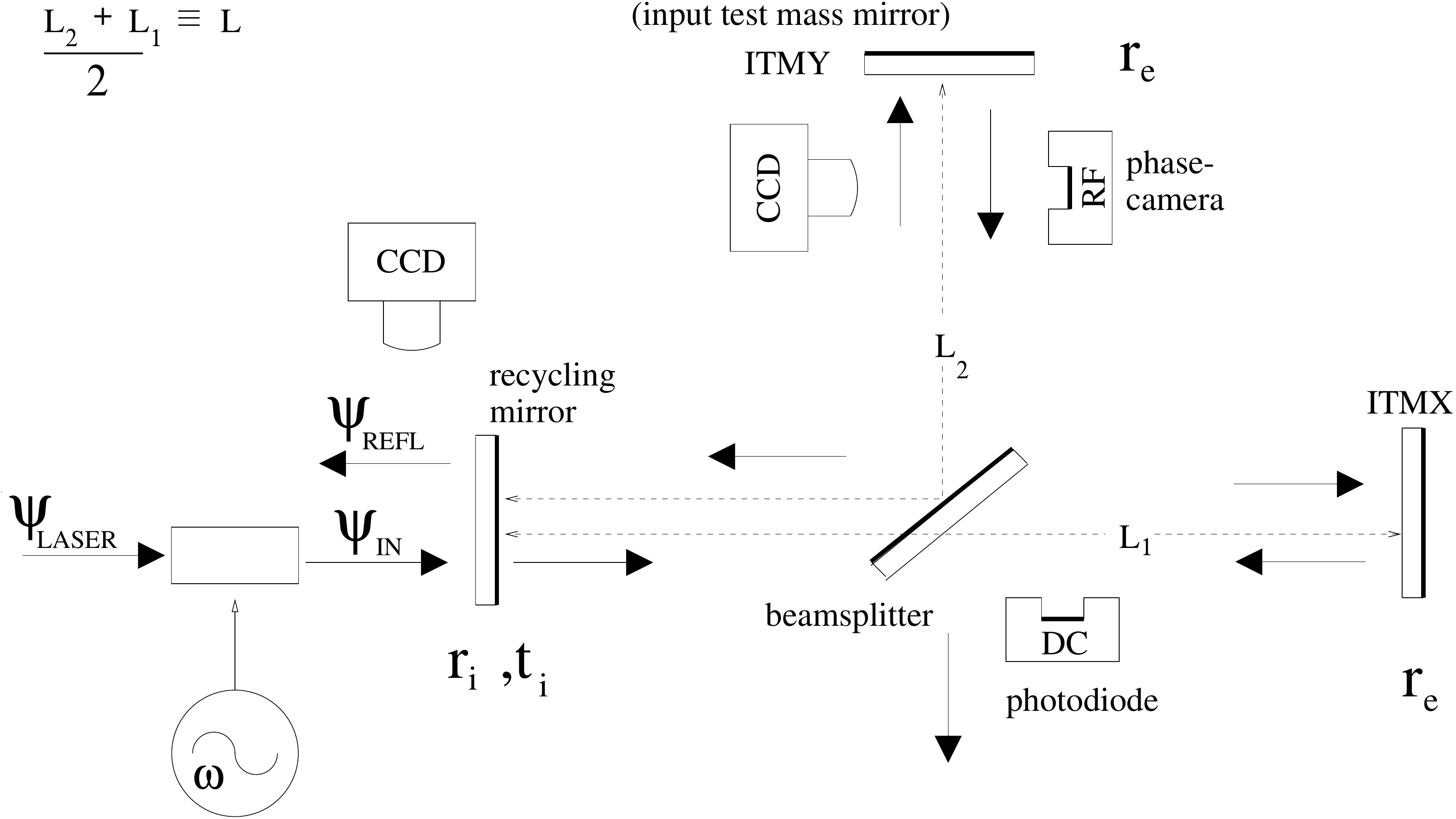}
    \caption{The laser field $\Psi_\scr{\mr{LASER}}$ is phase modulated to produce
    $\Psi_\scr{\mr{IN}}$ consisting (approximately) of carrier light and upper
    and lower phase modulation sidebands, 24.5~MHz on each side of the carrier. The
    modulation index is $\Gamma\approx0.34$.
    The reflected field $\Psi_\scr{\mr{REFL}}$ is the sum of three components: the sidebands
    (which are almost completely reflected back towards the laser since the
    cavity length makes them non-resonant), the prompt reflection of the carrier that
    is not interacting with the recycling cavity, and the leakage through the recycling
    mirror of the carrier field resonating inside the cavity. The cavity beam is
    split by the beam splitter into two optical paths of different lengths, represented in the
    diagram by $L_1$ and $L_2$. The amplitude transmission and
    reflection coefficients are represented by $t_i$ and $r_i$ respectively. The
    position of the beam splitter is actively controlled, so that its antisymmetric port
    corresponds to the dark fringe of the carrier.
    The various instruments are schematically arranged next to the beams they interrogate.
    Thus, one CCD camera interrogates the beam at the reflected port (REFL).
    Another CCD camera interrogates the beam from the y-leg of the
    recycling cavity picked off via the wedged side of ITMy. (This port is known as POY.)
    The RF phase camera interrogates this same beam also.
    The DC photodiode interrogates the beam from the x-leg of the
    recycling cavity picked off via the wedged side of the beamsplitter.
    (This port is known as POB).  Not shown is an RF photodiode registering the
    NSPOB signal (geometric mean of the upper and lower sideband powers)
    and located at the same port.  Like the phase camera, this photodiode was only
    used during the full interferometer lock discussed in \Secn{fulllock}.}
    \label{notation}
    \end{center}
    \end{figure}

The schematic diagram in \fig{notation} illustrates the general configuration and
introduces notation to be used later. To obtain information about the intensity
distribution of the fields interacting with the cavity, we used two CCD cameras: One
captured the light reflected from the cavity and the other captured light picked out of
the cavity by means of the slightly wedged antireflective side of one of the mirrors
(ITMy). We measured the power in the cavity using a calibrated DC photodiode to monitor
the intensity of the beam picked off at the slightly wedged, antireflective side of the
beamsplitter. The beam picked off at the antireflective side of the beamsplitter, like
the beam picked off at the antireflective side of the arm input test masses, is a good
representation of the recycling cavity beam.

In the case of the fully locked interferometer, additional methods were required to
provide information about the sidebands in the recycling cavity. When the full
interferometer is locked, the recycling cavity field intensity is dominated by the
carrier so we could not use a DC coupled photodiode or a CCD camera for the purpose of
interrogating the RF sidebands.  To measure the intensity of the sidebands, we used an RF
photodiode whose signal was demodulated at twice the sideband frequency. This photodiode
received a highly focused version of the beam picked off from the antireflective side of
the beamsplitter. The signal from this photodiode is called NSPOB and is proportional to
the geometric mean of the power in the upper and lower sidebands. We employed a similar
technique to measure the spatial profile of the sideband intensity using a
``phase~camera''~\cite{Goda2004}. The phase camera rapidly scans an enlarged version of
the beam over an RF photodiode which is very small compared to the size of the enlarged
beam. Like the NSPOB signal, the signal from this RF photodiode is demodulated at twice
the sideband frequency. However, due to the scanning, we now obtain a measurement of the
spatial profile of the geometric mean of the upper and lower sideband intensities. The
phase camera received the recycling cavity beam picked off from the slightly wedged
antireflective side of ITMy.

\section{FFT Model Overview}
The numerical simulation with which we compare our measurements---The FFT
Model~\cite{Bochner1998}---is based on a Fortran program whose first step is a Fourier
transformation of a grid representation of the optical field. In the wave vector domain,
a matrix multiplication provides the propagation of the field; then, when the interaction
with the optics must be reproduced, the optical field is Fourier transformed back from
the momentum representation to the spatial one and each field element on the grid is
multiplied by a phase delay describing the action of the optics.  Once iterative
propagation has terminated and the resulting fields are stationary, we construct the
Pound-Drever-Hall error signal directly from the fields.

\section{Results}
\subsection{Carrier resonant in the recycling cavity, arms non-resonant}
\label{crlock}

\begin{figure*}
    \includegraphics[width=6.4in]{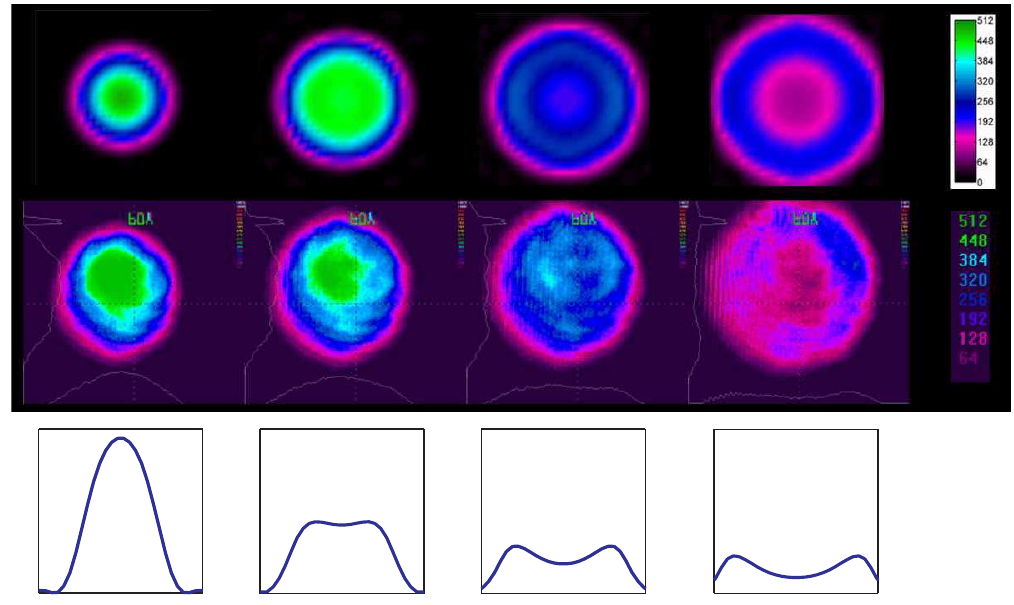}
    \caption{Effect of changing the cavity length by several nanometers on the shape of
    the carrier resonating in the recycling cavity. (Arms are not resonant.) Comparison
    of actual image captures with simulation results.
    The center row shows false color photographs of the recycling cavity beam picked off at ITMy for
    different cavity length offsets from the nominal lock point.
    From left to right, offsets are: {$-8$\1\1nm}, {$-4$\1\1nm}, {$0$\1\1nm}, and {$+4$\1\1nm}.
    The colors represent intensity and correspond to the linear scale shown at right (arbitrary units).
    The top and bottom rows show the FFT model results for the same cavity length offsets.
    The top row shows the simulation results for the beam intensity. Approximately the same color
    scale is used for the simulated intensity results as for the false color photographs of the
    center row so that the images can be directly compared. The bottom row shows the cross-sectional
    intensity from the simulation plotted as a function of radius. The units of distance represented by the
    axes are left arbitrary because the actual physical dimensions of
    images rendered by the camera were not recorded. (In other words, a camera calibration was not
    available.) However, the \emph{relative} sizes of the images in each individual row are accurate.
    A uniform zoom factor was applied to all the images in the top and bottom rows so that the size of the
    beam at {$0$\1\1nm} length offset (second from right) approximately matched the beam diameter
    in the corresponding image of the center row.}
    \label{CRlockdcPOY}
    \end{figure*}

\Fig{CRlockdcPOY} compares the resonant field intensity in the recycling cavity derived
from the FFT model with the actual measured field intensity. The four panels represent
the intensity distribution at four different cavity lengths, each 4~nanometers apart.
Clearly, the intensity distribution changes dramatically with very small changes of the
cavity length.

Note that the FFT model correctly reproduces the evolution of the beam shape as it goes
from a one-peak profile to a donut and also shows the increase in beam size as the offset
is changed between $-8$\1\1nm and $+4$\1nm, although the change in beam size is somewhat
less pronounced in the data than in the model.

\begin{figure}
    \begin{center}
    \includegraphics[width=3.2in]{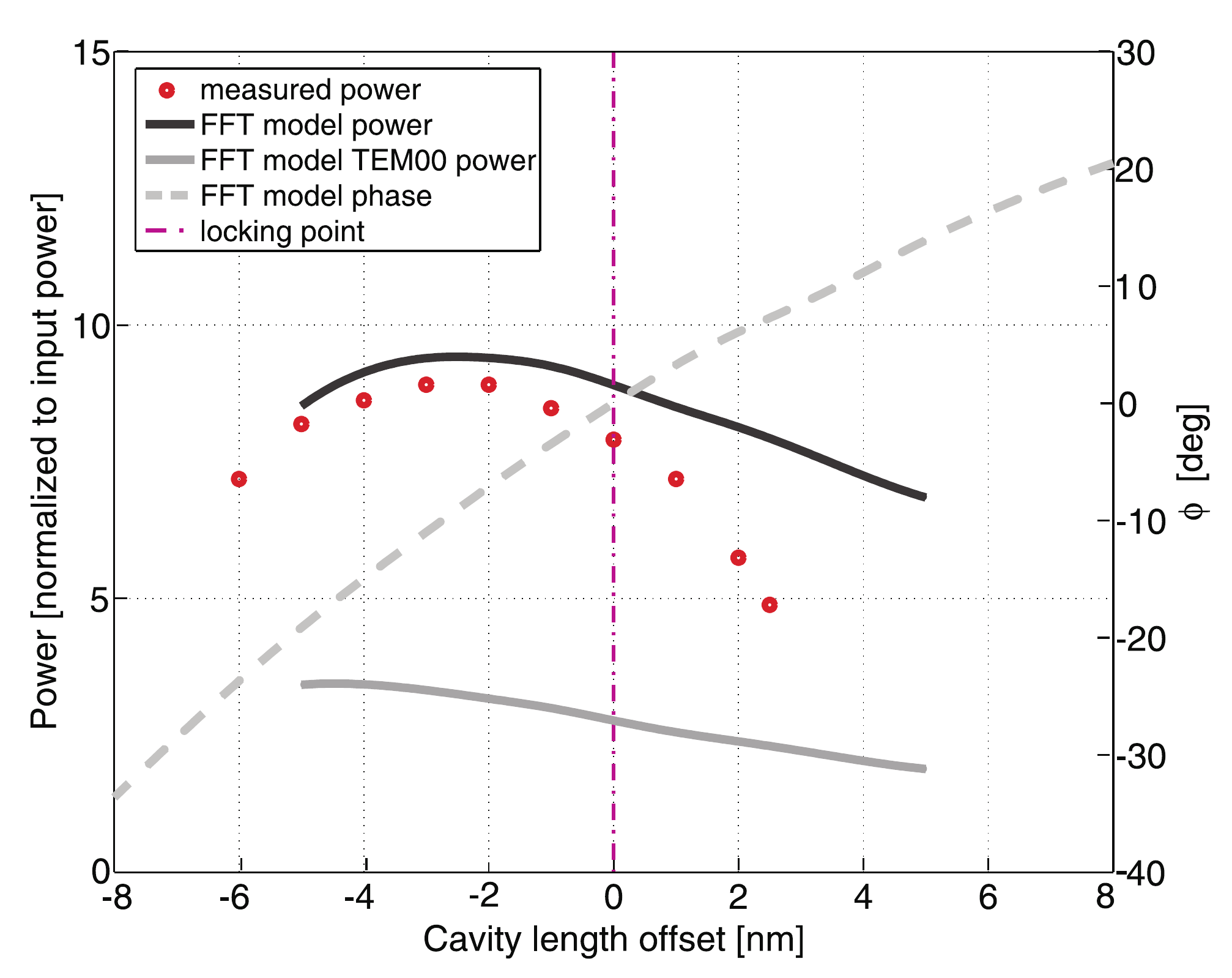}
    \caption{Power in the recycling cavity as a function of cavity length offset. The
    FFT Model correctly predicts the length offset at which the cavity power is
    maximized. The circles show the total cavity power measured at ten different length offsets.
    The dark solid line shows the FFT model prediction for the total cavity
    power. The light solid line shows the FFT model prediction for the power in the
    mode of the input beam to the cavity. The scale on the right refers to the
    light dashed line representing the FFT model phase
    $\phi=\mr{Arg}\left(\ip{\Psi_\scr{\mr{LASER}}}{\Psi_\scr{REFL}^\scr{{CR}}}\right)$ in the
    notation of \eqn{GeneralErrorSignal}.  $\phi$ is zero at the locking point.}
    \label{CRlockPPOBvsFFTmodel}
    \end{center}
    \end{figure}

\Fig{CRlockPPOBvsFFTmodel} compares the measured cavity power with the prediction from
the FFT model. The point of maximum power found by the FFT model agrees with the
measurement as does the approximate power build-up. The fact that the power data fall
generally below the prediction for the larger length offsets is likely due to a reduction
in the gain of the loop controlling the cavity length.  This produces a change in the
calibration leading to an underestimation of the cavity length change for those offsets
that are quite far from the point of maximum cavity power. The figure also shows the
portion of the power in the $\mr{TEM_{00}}$ mode, which is defined here as the mode of
the input beam to the recycling cavity (fundamental mode of the modecleaner propagated
through the mode-matching telescope towards the interferometer). By design, the input
beam mode is closely matched to the fundamental mode of the 4~km arm cavities.
Unexpectedly, and in contrast to the behavior seen in optically stable cavities (even
ones whose input beams are not matched to the cavity mode) the FFT model shows that the
power in the $\mr{TEM_{00}}$ component is maximized even further from the locking point
than the total cavity power.

To gain insight into this unusual behavior, we write down a general expression for the
Pound-Drever-Hall error signal that makes no assumptions about the spatial profile of the
beams involved. The notation refers to the fields indicated in \fig{notation}.  The
Pound-Drever-Hall error signal is generated by phase modulating the input beam to the
cavity.  The input field $\Psi_\scr{IN}$ can be expanded in terms of Bessel functions
\begin{equation}
\begin{array}{rcl}
    \Psi_\scr{IN} & = & \Psi_\scr{\mr{LASER}}\exp[i{\scriptstyle{\Gamma}}\cos{\omega}t] \\
    &\simeq&
    J_0({\scriptstyle{\Gamma}})\1\Psi_\scr{\mr{LASER}}+iJ_1({\scriptstyle{\Gamma}})
        \1\Psi_\scr{\mr{LASER}}\exp[i{\omega}t]\\
        &&+ iJ_1({\scriptstyle{\Gamma}})\1\Psi_\scr{\mr{LASER}}\exp[-i{\omega}t]+\ldots \\
    &\equiv& \Psi_\scr{IN}^\scr{{CR}}+\Psi_\scr{\mr{IN}}^\scr{SB+}\exp[i{\omega}t]+
        \Psi_\scr{\mr{IN}}^\scr{SB-}\exp[-i{\omega}t]+\ldots
\end{array}
\end{equation}
where $\w$ is the modulation frequency and $\Gamma$ is the modulation index. (In our
case, $\w=24.5~\mr{MHz}$ and $\Gamma\approx0.34$.) The input field is therefore often
considered as three collinear beams: the carrier, $\Psi_\scr{IN}^\scr{{CR}}$, and a pair
of sidebands, $\Psi_\scr{\mr{IN}}^\scr{SB+}$ and $\Psi_\scr{\mr{IN}}^\scr{SB-}$, with
frequency separation $\w$ on either side of the carrier frequency. The modulation
frequency is such that when the carrier beam is resonant in the recycling cavity (without
the arms resonant), the sidebands are nearly anti-resonant and thus nearly totally
reflected from the input mirror (the recycling mirror). In these circumstances, only the
carrier is significantly sensitive to the geometrical features of the cavity, including
length variations or alignment, as described by the equations
\begin{equation}
\begin{array}{rcl}
    \Psi_\scr{\mr{REFL}}^\scr{SB-}  &\simeq&-\Psi_\scr{\mr{IN}}^\scr{SB-} \\
    \Psi_\scr{\mr{REFL}}^\scr{SB+}  &\simeq&-\Psi_\scr{\mr{IN}}^\scr{SB+} \\
    \Psi_\scr{REFL}^\scr{{CR}}      &=& D\Psi_\scr{IN}^\scr{{CR}}
\end{array}
\end{equation}
where the operator $D$ incorporates the effect of the cavity on the resonant carrier. The
total reflected power $P_\scr{\mr{REFL}}$ is the integral of the squared field over an
area $S$ much larger than the beam size.
\begin{equation}\label{reflectedpower}
\begin{array}{rcl}
    P_\scr{\mr{REFL}}&=&\int_S
    |\Psi_\scr{\mr{REFL}}|^2\1dS\\
    &=&\int_S\left\{|\Psi^\scr{{CR}}_\scr{\mr{REFL}}|^2+
        |\Psi^\scr{SB+}_\scr{\mr{REFL}}|^2 + |\Psi^\scr{SB-}_\scr{\mr{REFL}}|^2  \right.\\
    &&  + 2\Re\left[(\Psi^\scr{{CR}}_\scr{\mr{REFL}}  {\Psi^\scr{SB-}_\scr{\mr{REFL}}}^* +
        \Psi^\scr{SB+}_\scr{\mr{REFL}}   {\Psi^\scr{CR}_\scr{\mr{REFL}}}^*)
        \exp(i{\omega}t)\right] \\
    &&  \left.+ 2\Re\left[\Psi^{\scr{SB+}}_{\scr{\mr{REFL}}}
        {\Psi^{\scr{SB-}}_{\scr{\mr{REFL}}}}^*\exp(2i{\omega}t)\right]+\ldots \right\}dS\,.
\end{array}
\end{equation}

We are interested in the error signal for the length of the cavity. This error signal,
which controls the position of the recycling mirror, is basically the same as the error
signal from a simple two-mirror cavity. The error signal, which we call $V_I$ here, is
the cosine-phase of the demodulated voltage from an RF photodiode placed at the reflected
port sensing $P_\scr{\mr{REFL}}$. The beam is focused onto the active area of the RF
photodiode so that no beam clipping occurs. Using the modulation frequency $\omega$ to
demodulate the signal from this photodiode, we find that one term survives.
\begin{equation}
\begin{array}{rcl}
    V_I\,&\approx& \alpha\int_S
    dS\int_{0}^{T}\!dt\,\frac{P_{\scr{\mr{REFL}}}\cos({\omega}t)}{T}\\
    &=&\alpha\int_S\Re\1(\Psi^\scr{{CR}}_\scr{\mr{REFL}}
    {\Psi^\scr{SB-}_\scr{\mr{REFL}}}^* +
        \Psi^\scr{SB+}_\scr{\mr{REFL}}   {\Psi^\scr{CR}_\scr{\mr{REFL}}}^*)\1dS \\
\end{array}
\end{equation}
where $\Re$ indicates the real part, $T \gg \omega^{-1}$ is the effective integration
time of the sensing chain and $\alpha$ is an overall constant representing the efficiency
of the photo-detection and the gain of the sensing chain. Rewriting the integrals as
inner products brings out the structure\1---
\begin{equation}\label{GeneralErrorSignal}
\begin{array}{rcl}
     V_I &\propto& \Re\left[\,
        \ip{\Psi^\scr{{CR}}_\scr{\mr{REFL}}}{\Psi^\scr{SB-}_\scr{\mr{REFL}}}
        + \ip{\Psi^\scr{SB+}_\scr{\mr{REFL}}}{\Psi^\scr{CR}_\scr{\mr{REFL}}}
        \,\right] \\
        &=& 2 J_1({\scriptstyle{\Gamma}})
        \;\Im\ip{\Psi_\scr{\mr{LASER}}}{\Psi_\scr{REFL}^\scr{{CR}}}
\end{array}
\end{equation}
where $\Im$ indicates the imaginary part. \Eqn{GeneralErrorSignal} makes it obvious that
the error signal is generated only from the component of the returning carrier that is in
the mode of the input beam.

In \emph{stable} cavities, different spatial modes of the cavity are separated by a
discrete Gouy phase and therefore resonate at slightly different cavity lengths.  The
Pound-Drever-Hall servo generates a large error signal whenever the cavity field has
large overlap with the mode of the input beam.  Using mode-matching optics, we normally
arrange for large overlap to occur with only one of the cavity modes (usually the
fundamental mode). The Pound-Drever-Hall servo will lock the cavity onto the chosen mode
because this is the only mode contributing significantly to the error signal. When the
cavity locks  on a low-loss mode to which the input beam is well matched, large buildups
of that cavity mode can occur. In that situation, the imaginary part of the overlap
integral in \eqn{GeneralErrorSignal} becomes zero precisely when the phase of the carrier
component exiting the cavity through the input mirror matches the phase of the input beam
and is therefore 180\degrees out of phase with the promptly reflected beam. This
maximizes the power of the component of the cavity field that is in the mode of the input
beam, because this component of the cavity field experiences a destructive phase
condition with the promptly reflected beam. In short, when $V_I$ is zero, we get maximum
power in that component of the cavity field \emph{that is in the mode of the input beam}.
For stable cavities, where the shape of the cavity mode does not change with small length
changes of the cavity, $V_I=0$ must therefore also correspond to the point of maximum
cavity power, regardless of mode.

For marginally unstable cavities $(g_1g_2\rightarrow 1)$ the eigenmode decomposition
breaks down and we observe that the transverse shape of the cavity beam depends strongly
on the cavity length. The Pound-Drever-Hall error signal is modified by this spatial
dependence in such a way that minimization of the error signal, and therefore the natural
lock point, no longer corresponds to maximizing the input beam mode component of the
cavity field. Nor is there any particular reason to expect that the overall cavity power
is maximized at the natural lock point. From the naive point of view therefore, the servo
has developed an offset. As discussed above, this effect was clearly seen in the LLO
recycling cavity data and in the model.

The dependence of the cavity beam shape on the cavity length is probably due to two
effects. First, as the fields inside the cavity propagate, they spread out slightly due
to diffraction. Thus, fields corresponding to consecutive cavity traversals do not have
exactly the same spatial beam profile. With length changes alone, it's therefore not
possible to enforce constructive superposition everywhere in the cavity between beams
that have traversed the cavity a different number of times. Thus, beam propagation within
the cavity is bound to lead to beam shaping, simply due to the fact that some regions
will have constructive superposition while others have destructive superposition. In
addition, this shape can be expected to depend strongly on the cavity length since even a
small change in the cavity length alters the interference condition between cavity
traversals. Secondly, the shape and location of regions of destructive phase between the
cavity beam and the promptly reflected beam at the input mirror depends on the intensity
and phase profile of the cavity beam. Thus, the shape and location of those regions at
the input mirror where light is efficiently coupled into the cavity also change with
microscopic cavity length changes. Under these conditions, the power of the input beam
mode in the cavity is set by a combination of direct coupling of the input beam mode into
the cavity (requiring destructive interference with the promptly reflected beam) and the
transferral of power coupled into the cavity in a combination of modes into the mode of
the input beam.  Since destructive interference of the input beam mode in the cavity with
the promptly reflected beam corresponds to $V_I=0$, we should not be surprised to find
that this alone does not lead to maximum input beam mode power in the cavity.  And of
course there is no obvious reason to expect the overall cavity power to be maximized for
zero error signal either. Indeed, we have no particular reason to expect that the cavity
length offset corresponding to maximum overall cavity power corresponds to the point of
maximum power in the input beam mode component of cavity field, as illustrated by the FFT
model results.

Qualitatively consistent behavior has been observed at LIGO Hanford Observatory (LHO) in
the marginally unstable recycling cavities of their
interferometers~\cite{Kells2004,Ottaway_Private}. As in the LLO interferometer, the
maximum cavity power build up occurs only when an offset is applied to the natural lock
point. Transverse beam profile changes are evident as well. (In fact, these effects were
first seen at LHO, and subsequently at LLO, several years before the current study to
quantitatively compare the FFT Model predictions with the LLO behavior was begun.)

The high level of agreement between the experimental observations and the predictions of
the FFT Model, indicate that the observed loop offset is a true optical effect (not due
to a simple technical problem such as unintended offsets in the control electronics).
Such an offset can therefore be expected to develop in any marginally unstable cavity
locked by the Pound-Drever-Hall technique and the resonant points of such cavities may
need to be adjusted ``by hand'' to compensate.

\subsection{Full interferometer lock}
\label{fulllock} When the full interferometer is locked~\cite{RegehrThesis}, the
intensity in the recycling cavity is dominated by the carrier whose spatial structure is
set by the input conditions to the arm to be the $\mr{TEM_{00}}$ mode of the optically
stable arm cavities.  This is due to the fact that the arms are overcoupled.  Thus, the
total carrier field (promptly reflected field plus leakage field) returning from the arms
is 180 degrees out of phase with the promptly reflected field alone and has approximately
the same magnitude as the incident field. As a result, the recycling cavity length
leading to resonance of carrier light that is also resonant in the arms is different by
one half wavelength than the length leading to resonance of carrier light that is not
resonant in the arms. In other words, higher order carrier modes (in the basis of the
arms) are anti-resonant in the recycling cavity while the $\mr{TEM_{00}}$ mode of the
arms is resonant.  Thus, the carrier field in the recycling cavity is entirely in the
$\mr{TEM_{00}}$ mode of the arms.  Now, the frequency of the sidebands was chosen such
that they would be resonant in the recycling cavity \emph{without} resonating in the arms
precisely when the carrier that \emph{does} resonate arms is resonant in the recycling
cavity.  Thus, the recycling cavity light is a mixture of $\mr{TEM_{00}}$ mode carrier
light and sideband light in a very large number of modes, with the carrier light
dominating the intensity due to the greater power in the carrier.

\begin{figure}
    \includegraphics[width=3.2in]{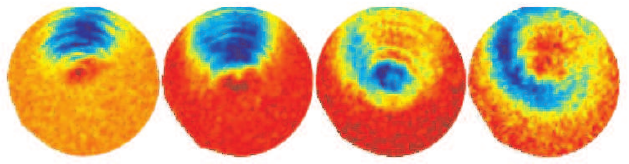}
    \caption{Effect of changing the cavity length by several nanometers on the shape
    of the sidebands resonating in the recycling cavity. The colors represent the
    geometric mean of the intensity of the upper and lower sidebands.  Blue corresponds to
    the regions of greatest intensity with orange/red corresponding to the regions of lowest
    intensity.
    The second image from right represents the natural lock point (zero applied offset).
    Note that in these images, the center of the beam is in the upper half of the image.
    The asymmetric structure of the beam in some of the images, particularly the two at
    right, is due to pitch and yaw motion of the optics to which the instantaneous field
    distribution is very sensitive.}
    \label{FullLockPhasecamera}
    \end{figure}

The sidebands, being non-resonant in the arms experience almost identical conditions as
the carrier in \secn{crlock} (where the carrier is resonant in the recycling cavity with
the arms non-resonant). Therefore, we can expect the field structure of the sidebands in
the full lock to be qualitatively similar to the carrier field distribution of
\secn{crlock}. The input power to the recycling cavity during these measurements (roughly
one Watt) was insufficient to generate significant thermal lensing in the recycling
cavity optics, even when the interferometer was fully locked. As before, the thermal
compensation system was not operating. Phase camera images of the sideband intensity
taken from the beam picked off at the antireflective side of ITMy are shown in
\fig{FullLockPhasecamera} at four different recylcing cavity length loop offsets. For
technical reasons, we did not obtain a calibration of the length loop offset in terms of
cavity length change.

\begin{figure}
    \begin{center}
    \includegraphics[width=3.2in]{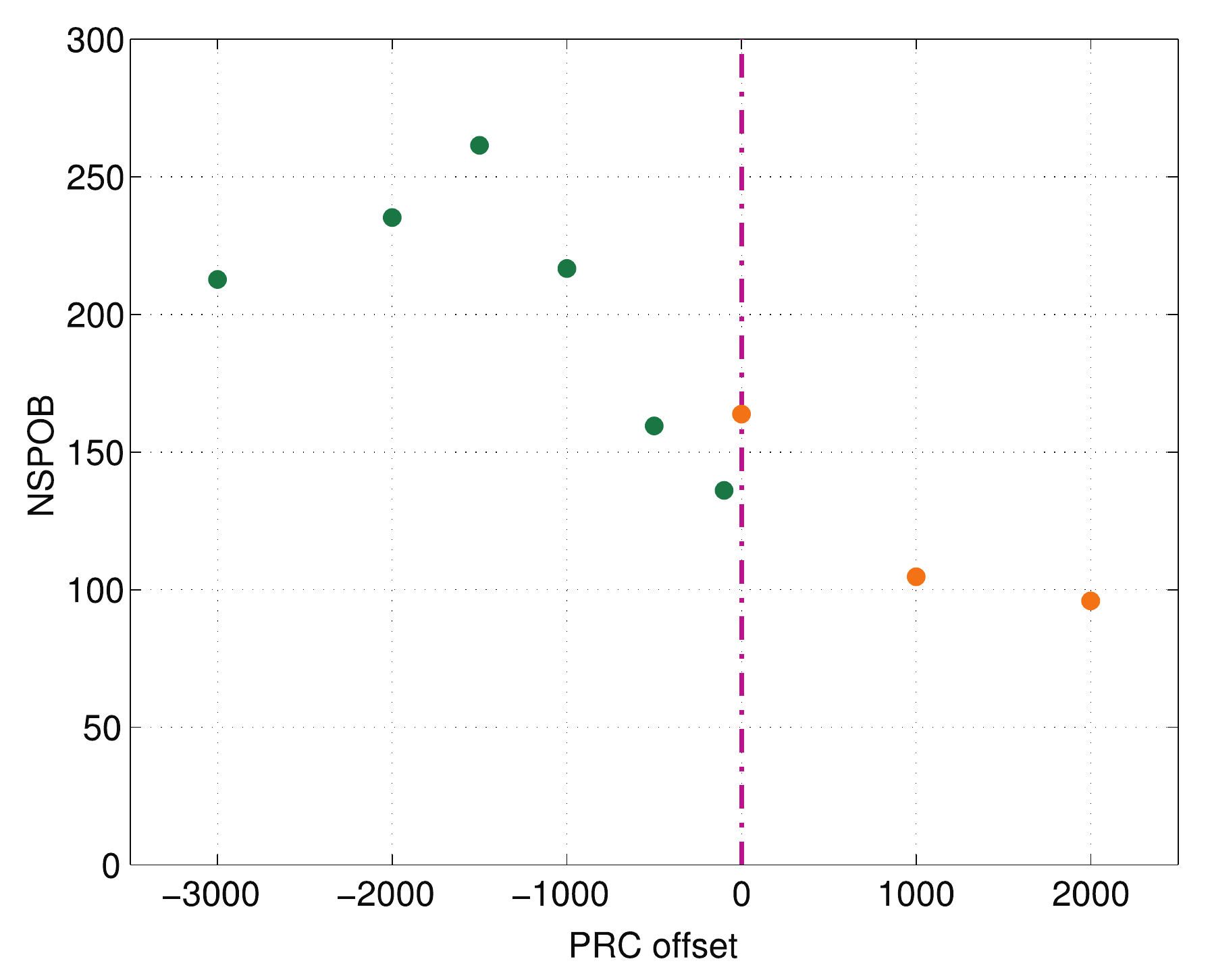}
    \caption{Sideband intensity as a function of uncalibrated length offset. This
    graph contains data from two lock stretches which we have distinguished by the
    two different hues.}
    \label{FullLockSPOB}
    \end{center}
    \end{figure}

\Fig{FullLockSPOB} shows the (geometric mean of the) power of the sidebands in the
recycling cavity as the cavity length is changed. The x-axis shows the offset added to
the error point in uncalibrated counts. The y-axis is the value of the interferometer
NSPOB signal, which measures the geometric mean of the upper and lower sideband powers.
Note that we see a much larger change in the sideband intensity in this fully locked
state than the change in the carrier intensity for the carrier lock in the recycling
cavity. The reason for this may be that as the profile of the sidebands in the recycling
cavity begins to better match the resonating mode of the arm cavities, a larger fraction
of the beam is exactly on anti-resonance and is totally reflected from the arm thus
enhancing the build up of the sidebands in the recycling cavity.

As expected, the qualitative behavior of the sideband power and the qualitative
transverse profile changes of the sidebands are the same as those seen in \secn{crlock}.
As before, the shape of the cavity beam changes as a function of small length changes of
the cavity, and the lock point does not correspond to maximum sideband power in the
cavity. Although we could not extract the $\mr{TEM_{00}}$ component of the sidebands from
the intensity alone, it seems likely based on the analysis of \secn{crlock} that it is
also not maximized.

\section*{Acknowledgements}

The continuing and reliable operation of the  LIGO interferometer at Livingston was made
possible by the hard work of our many commissioning colleagues.  Rana Adhikari taught one
of the authors [AMG] how to understand the LIGO Livingston interferometer and also
provided useful comments for the paper. David Ottaway read versions of the manuscript and
shared insights based on his considerable knowledge of the interferometer optics.  He was
also one of the first people to appreciate the importance of the non-gaussian structure
of the sidebands in the LLO recycling cavity, first seen during installation and
commissioning of the LLO phase camera with one of the authors [AMG].  Bill Kells and
David Ottaway provided information about the LHO results and Bill Kells pioneered the
application of the Eikonal optical model. The authors thank Riccardo DeSalvo and David
Reitze for encouraging the submission of the manuscript for publication. Finally, Keita
Kawabe provided a careful internal (LSC) review of the manuscript and made observations
leading to improvements. LIGO was constructed by the California Institute of Technology
and Massachusetts Institute of Technology with funding from the National Science
Foundation and operates under cooperative agreement PHY-0107417. This paper has LIGO
Document Number LIGO-P070044-02-Z .

\bibliographystyle{apsrev}

\end{document}